\documentclass[12pt,a4paper]{llncs}
\usepackage{fullpage}

\newcommand{\Oh}[1]
  {\ensuremath{\mathcal{O}\! \left( {#1} \right)}}
\newcommand{\SA}
  {\ensuremath{\mathsf{SA}}}

\begin{document}

\title{Suffix Arrays for\\Spaced-SNP Databases}
\author{Travis Gagie\thanks{Supported by the Academy of Finland.}}
\institute{Department of Computer Science\\University of Helsinki, Finland}
\maketitle

\begin{abstract}
Single-nucleotide polymorphisms (SNPs) account for most variations between human genomes.  We show how, if the genomes in a database differ only by a reasonable number of SNPs and the substrings between those SNPs are unique, then we can store a fast compressed suffix array for that database.
\end{abstract}

\section{Introduction}
\label{sec:introduction}

Indexing large genomic databases is a challenging problem in bioinformatics, and several authors have designed heavily-engineered data structures specifically for this purpose; see, e.g.,~\cite{DDG14} and references therein.  In this paper we propose a simple model of these databases and give a theoretical solution, which we hope will give insight into the real problem.

Specifically, we assume that the genomes in the database differ only by a relatively small number of single-nucleotide polymorphisms (SNPs) --- i.e., single-character substitutions that occur in the DNA of a significant fraction of the population --- and that the substrings between the SNP sites are unique.  That is, we do not consider insertions, deletions or SNPs that are not uniquely distinguished by the fixed substring that follows them.  We feel our model is a reasonable first approximation since there are a few million SNPs in the human genome, averaging one every few thousand base-pairs, and they account for most of our genetic variation.  For simplicity, we consider only two-allele SNPs since they are the most common.

We show how, under these assumptions, we can store the suffix array for a database of $m$ genomes of length $n$ with $k$ SNP sites in $\Oh{n + k m / \sqrt{\log n}}$ space and support access to the SA in $\Oh{1}$ time.

\section{SSNP Databases}
\label{sec:definition}

Consider the language $L$ accepted by a regular expression
\[R = \alpha_1 \circ \Sigma \circ \alpha_2 \circ (a_2 + a_2') \circ \cdots \circ \alpha_k \circ (a_k + a_k') \circ \alpha_{k + 1}\,,\]
where \(\alpha_1, \ldots, \alpha_{k + 1}\) are non-empty substrings, \(a_1, a_1', \ldots, a_k, a_k'\) are characters, $\circ$ indicates concatenation and + indicates disjuction.  We call $L$ a $k$-site spaced-SNP ($k$-SSNP) language if no word in $L$ can contain more than one occurrence of each of \(\alpha_1, \ldots, \alpha_{k + 1}\).

Assume $L$ is a $k$-SSNP language, let \(w_1, \ldots, w_m \in L\) and let
\[D = w_1 \circ \# \circ w_2 \circ \# \circ \cdots \circ \# \circ w_m \circ \#\,,\]
where $\#$ is a special character lexicographically strictly less than any character in the alpabet of $L$.  We call the string $D$ a $k$-SSNP database; notice it has length \(m (n + 1)\), where \(n = |w_1| = \cdots = |w_m|\), but it can be represented by $R$ and an \(m \times k\) binary matrix (which may also be compressible).

\section{Representing Suffix Arrays}
\label{sec:representation}

Let $L$, $R$ and $D$ be as described in the previous section.  The suffix array (SA)~\cite{MM93} of $D$ is the array \(\SA [1..|D|]\) in which \(\SA [i]\) stores the starting position of $D$'s lexicographically $i$th suffix.  Although storing $\SA$ explicitly takes \(\Theta (|D|)\) space, we can use the properties of SSNP databases to compress it.

\subsection{Blocking}
\label{subsec:blocking}

First of all, since $\#$ is lexicographically less than any character in the alphabet of $L$,
\[\SA [1..m] = [m (n + 1), (m - 1) (n + 1), \ldots, 2 (n + 1), n + 1]\,.\]
Now consider $D$ as a matrix with $m$ rows containing \(w_1, \ldots, w_m\), and \(n + 1\) columns.  Choose a column $c$ and let $c'$ be the next column containing two distinct characters (i.e., the next SNP site) or $\#$.

Let \(r_1, \ldots, r_t\) be the rows that contain one of the two distinct characters in column $c'$.  Since the substring $\alpha$ following the $c'$th character in every word in $D$ is unique in that word, the values in
\[C = \{c + (r_1 - 1) (n + 1), c + (r_2 - 1) (n + 1), \ldots, c + (r_t - 1) (n + 1)\}\]
appear together in $\SA$ and, moreover, their order of appearance there is the same as the order of appearance of the values in
\[C' = \{c' + (r_1 - 1) (n + 1), c' + (r_2 - 1) (n + 1), \ldots, c' + (r_t - 1) (n + 1)\}\,,\]
which also appear together in $\SA$.

Suppose we have explicitly stored the values in $C'$ where they occur in $\SA$.  Instead of storing the values in $C$ explicitly, we store only \(c' - c\) and a pointer from the block of $\SA$ that would contain them to the block of $\SA$ containing the values in $C'$.  Doing this for all the columns takes $\Oh{n + k m}$ space.  We also store a bitvector with 1s marking the beginnings of blocks of $\SA$ for which we do not have values explicitly stored, which takes $\Oh{n}$ bits.

To access a value in $\SA$ that we do not have explicitly stored, we use rank and select on the bitvector to determine which block that value is in and its offset in that block; follow the block's pointer to a block of explicitly stored characters; find the corresponding explicitly stored value; and subtract the difference between the appropriate columns.

\subsection{Permutations}
\label{subsec:permutations}

We can save even more space if we take advantage of the fact that each SNP changes the lexicographic order of the suffixes in a simple way.  Let $c'$ and \(r_1, \ldots, r_t\) be as described in the previous subsection and let $c''$ be the next column containing two distinct characters or $\#$.

Since
\[D [c' + (r_i - 1) (n + 1)..c'' + (r_i - 1) (n + 1) - 1]
= D [c' + (r_j - 1) (n + 1)..c'' + (r_j - 1) (n + 1) - 1]\]
for \(i, j \leq t\), the order of appearance of \(c' + (r_i - 1) (n + 1)\) and \(c' + (r_j - 1) (n + 1)\) in $\SA$ is the same as the order of appearance of \(c'' + (r_1 - 1) (n + 1)\) and \(c'' + (r_1 - 1) (n + 1)\).

It follows that, once we have the order of appearance of
\[\{c'' + n + 1, c'' + 2 (n + 1), \ldots, c'' + m (n + 1)\}\,,\]
we can store the order of appearance of
\[\{c' + n + 1, c' + 2 (n + 1), \ldots, c' + m (n + 1)\}\,,\]
using only a bitvector of length $m$.  That is, the permutation that changes the former order into the latter one can be partitioned into two incrementing subsequences.  We can store the composition of $p$ such permutations using \(m p\) bits such that evaluating the permutation takes $\Oh{1}$ time~\cite{GMV14}, using data structures for access, partial rank and select on a string in \(\{0, \ldots, 2^p - 1\}^m\).

If we store explicitly the blocks for every $\sqrt{\log n}$-th SNP-site column, then we use $\Oh{n + k m / \sqrt{\log n}}$ space and can still access $\SA$ in $\Oh{1}$ time.  We should use still less space when the distribution of characters in the SNP-site columns is skewed.

\end{document}